\documentclass[preprint]{aastex}

\usepackage{graphicx}

\usepackage{natbib}

\newcommand{\be}{\begin{equation}}
\newcommand{\ee}{\end{equation}}
\newcommand{\nn}{\mbox{} \nonumber \\ \mbox{} }
\newcommand{\ba}{\begin{eqnarray}}
\newcommand{\ea}{\end{eqnarray}}
\newcommand{\om}{\omega}

\newcommand{\Bf}{{magnetic field\,}}

\newcommand\etal{\textit{et al.\ }}
\newcommand\eg{\textit{e.g.\ }}

\begin{document}
\title{Orbital modulation of emission of the binary pulsar J0737-3039B}
\author{
Maxim  Lyutikov}
\affil{University of British Columbia, 6224 Agricultural Road,
Vancouver, BC, V6T 1Z1, Canada}

\begin{abstract}
In  binary radio pulsar system J0737$-$3039,  slow pulsar B
shows orbital  modulations of intensity, being especially bright at two 
short orbital  phases. We propose  that  these modulations are due to 
distortion of pulsar B magnetosphere 
by pulsar A wind which produces orbital phase-dependent changes
of the    direction 
along which radio waves are emitted. In our model,  pulsar B is intrinsically
bright at all times but its radiation  beam misses the Earth
at most  orbital phases.
We employ a simple model of  distorted B magnetosphere using stretching
transformations of Euler potentials  of  dipolar fields.
To fit observations we use  parameters of  pulsar B derived from 
  modeling of A eclipses (Lyutikov \& Thompson 2005).
The model reproduces two bright regions approximately at
the observed orbital phases,  explains
 variations of the pulse shape between them  and
regular timing residuals within each emission window. It also
makes predictions for timing properties  
and secular  variations of pulsar B profiles.
\end{abstract}

\section{Introduction}

Double pulsar system PSR J0737$-$3039A/B contains a recycled 22.7
ms  pulsar (A) in a 2.4 hr orbit around a 2.77 s pulsar (B)
\citep{burgay03,lyne}. 
Emission of pulsar B is strongly dependent on  orbital phase.
It is especially bright at two windows,  each lasting for about $30^\circ$.
 One of the  emission windows
appears near superior conjunction 
(when pulsar B is between pulsar A and observer), and another approximately 
$70^\circ$ before it.
Pulse profiles have different shapes in the two windows. 
In addition, \cite{ransom} have detected 
 regular, orbital phase-dependent
drift of emission arrival times by as much as 20 ms.

Previously,
\cite{jenet04} suggested that emission of B is initiated by  $\gamma$-ray
emission from A.
This  model
seems to be inconsistent with the evolution of the A profile \citep{manch05}.
\cite{zl04} suggested that  B emission is triggered by particles from pulsar A
wind reaching deep into magnetosphere. This model  is incorrect
since the authors neglected magnetic bottling effect,
which would  reflect most pulsar A wind  particles high above in B magnetosphere 
\citep[see][for discussion of particle dynamics inside B magnetosphere]{lt05}

In this paper we explore an alternative possibility   that  orbital brighting 
of B is due to distortions of  B magnetosphere by pulsar A wind. We show that due to
orbital phase-dependent
distortions of B magnetosphere,  the polar 
magnetic field lines, along which emission is presumably generated, may be ``pushed''
in the direction of 
an observer at particular orbital phases, while  at 
other moments radiating beam of B
misses the observer.

\section{Distortion of  pulsar B magnetosphere}

Magnetosphere of pulsar B is  truncated, if compared  to
magnetosphere of an isolated pulsar of the same spin,
by the relativistic wind flowing outward from pulsar A.
The  size of  magnetosphere is $R_m \sim 4 \times 10^9$ cm \citep{lyut04,arons,lt05},
which is 3 times smaller than the 
light cylinder radius $R_{LC}=\Omega/c=  1.3 \times 10^{10}$ cm
 ($\Omega$ is angular frequency of pulsar B rotation, $c$ is velocity of light).
At intermediate distances, $  R_{NS} \leq r \leq ( R_{LC}, \, R_m)$,
neutron star magnetospheres
are well  approximated by  dipolar structure. 
This has been a longstanding assumption in pulsar theory  \citep{gj69}, 
and has recently been confirmed by modeling of pulsar A eclipses
\citep{lt05}. 

A
small degree of distortion of magnetosphere from the dipolar shape
is expected at intermediate distances 
  due to   several types of electrical currents.
First, distortions are due to confining Chapman-Ferraro currents
\citep{chapfer}
flowing in the magnetopause (a region of  shocked pulsar A wind around B magnetosphere).
At the emission radius, $R_{em} \sim 1-5 \times 10^8$ cm (see below),
fractional
 distortions due to  Chapman-Ferraro currents
 are expected to be small $\sim (R_{em} /R_{m})^3 \sim .001-.02$.
In addition to confinement  of magnetosphere on the side facing pulsar A (``dayside''),
on the side opposite to pulsar A (``nightside'') magnetosphere of B extends
to large distances, 
somewhat similar to the Earth magnetosphere under
the influence of Solar wind. 
 Secondly, similar to 
 isolated pulsars, there 
are conduction  Goldreich-Julian currents, arising on the open field lines
due to relativistic electromagnetic   effects of rotation,
 and  displacement currents,  arising in oblique rotators
due to temporal variations of electromagnetic  fields.
At the emission radius  these currents
produce a distortion of magnetic field of the order $\sim (R_{em} /R_{LC})^2 \sim 10^{-3}$.
Thirdly, there are internal currents flowing  in the magnetosphere, 
like ring current, Birkland currents, and   other types of currents.

Clearly, the detailed structure of pulsar B magnetosphere 
is even more complicated than that of the Earth, but at 
 distances from the star smaller than 
light cylinder and magnetospheric radii  distortions from
dipolar form are expected to be small.
This should be the case in the emission generation region.
In addition, since magnetospheric 
radius  $R_m$ is several times smaller than 
light cylinder we can make a simplifying 
assumption  that
at each given moment the structure of the inner
magnetosphere  is determined by the instantaneous direction of the magnetic moment of
B and the direction of line connecting two pulsars 
(which is, approximately, the direction of pulsar A wind at the position of B). 
Under this approximation we expect that at each moment the structure of the
inner magnetosphere may be estimated using methods developed for 
non-relativistic, quasi-stationary magnetospheres
of solar planets interacting with
the solar wind. This interaction is complicated, depending on
a number of both macroscopic (\eg wind pressure, direction of magnetic field,
dipole inclination) and especially microscopic (\eg reconnection and diffusion
rates) parameters. 
In what follows we use experience with the solar wind -- Earth magnetosphere interaction 
\citep[\eg][]{tsyg90}
as a guiding  line in studying 
 pulsar B magnetosphere. 

One of the principal issues here is how  magnetopause currents 
 respond to the  dipole magnetic field of the central object. 
Two extreme possibilities are (i) complete screening of the  dipole, so that
no pulsar magnetic field lines penetrate into the wind and (ii) efficient reconnection
so that most of the pulsar magnetic field lines that  reach the magnetospheric boundary 
penetrate into the wind. In the case of  the Earth magnetosphere, though reconnection
plays an important role, on average inter-penetration  of magnetic field is at most
a $10\%$ effect \cite[eg.][]{Stern87}. 
(Rates of reconnection
are strongly dependent  on the direction of the solar wind magnetic field.
On the dayside 
reconnection  occurs most efficiently near  the cusps, where polar magnetic field lines
reach magnetopause.)
 Numerical modeling of interaction of relativistic pulsar A wind with
pulsar B shows  qualitatively  similar results \cite{arons}.
Thus, as a first approximation, we may assume that 
magnetopause currents screen out pulsar B  magnetic field.

An additional source of magnetic field distortion 
is the ring current generated by  particles  trapped
inside magnetosphere.
Modeling of pulsar A eclipses \citep{lt05} 
implies that high density, relativistic
plasma (most likely composed of pairs) is present on closed field lines 
of pulsar B.
In addition to bouncing between magnetic poles due to effect 
of magnetic bottling, trapped particles drift along magnetic equator.
Viewed from the north magnetic pole, positrons drift clockwise, electrons --
counterclockwise,  producing  a  ring current, which 
modifies the structure of
 magnetosphere. We expect that effects of the  ring current are negligible 
at the emission radius. A typical drift 
velocity of charge carries is $u_d\sim c^2 \gamma_0 /\om_B R_m 
\sim 3 \times 10^4 $ 
cm/sec 
(here $\gamma_0 \sim 10$ is a typical Lorentz factor of trapped particles,
$\om_B$ is cyclotron frequency).
For particle density $n\sim \lambda _{m} n_{GJ,m}$ ($ n_{GJ,m}$ is Goldreich-Julian density at the magnetospheric radius $R_m = 4 \times 10^9$ cm
and $\lambda _{m}$ is multiplicity factor at $R_m $),
the current density is
$j\sim \lambda _{m} n_{GJ,m} u_d e \sim .1 j_{GJ}$, and 
total current is  $I \sim j R_m^2$. Resulting magnetic field is
$B_{ring} \sim I/(R_m c) \sim j R_m /c \sim \lambda _{m}
( \Omega R_m /c) (u_d/c) B_m \sim .03  B_m $.
Deep inside magnetosphere, the  magnetic field of the ring current is nearly constant,
 but the dipole field increases,  so that at $R_{em} \sim 10^8$ cm
$ B_{ring} / B_d \sim 10^{-5}$. (Qualitatively, the  \Bf of the  ring current
would become comparable to the dipole field at $R_m$ when energy
density of trapped plasma is of the order of \Bf energy density.)

There are other types of currents that can modify field structure like Birkland and
tail currents \cite[\eg][]{tsyg90}. Their influence on the structure of magnetosphere at intermediate
distances is expected to be small and we neglect them here.
Thus,  we assume that the only  currents
contributing to the distortion of  
magnetosphere are magnetopause Chapman-Ferraro currents
which screen pulsar B field. This simplification allows us to 
use models
developed for the Earth magnetosphere in order to  
estimate distortions of pulsar B magnetic field.

\section{Modeling distorted magnetosphere of B}

There is extensive
literature on modeling of the Earth  magnetic field \citep[\eg][]{tsyg90}. 
A number of analytical methods have been developed. 
For our purposes, we do not need to calculate  the full structure of 
magnetosphere, but only to estimate variations in the position of  polar 
magnetic field lines, where emission of B is presumably generated.
For this purpose we employ the method of distortion transformation of 
Euler potentials \citep{Stern94,Voigt}. 
A major advantage of the stretching model of magnetosphere  is that it reproduces fairly 
well the structure of  a
tilted dipole \citep{Stern94}.

Magnetic field can  be  described by two 
Euler potentials $\alpha$ and $\beta$ (sometimes  called Clebsch potentials):
\be
{\bf B} = \nabla \alpha \times  \nabla \beta
\ee
so that magnetic field line is defined by an
intersection of  surfaces with constant 
 $\alpha$ and $\beta$.
Magnetosphere  of B enshrouded by   magnetopause resembles
a dipole field compressed on the dayside and stretched out on the nightside. The structure
of the nightside 
magnetosphere 
can be approximated  by stretching transformations of the 
Euler potentials $\alpha$ and $\beta$.

\begin{figure}[h]
\includegraphics[width=0.9\linewidth]{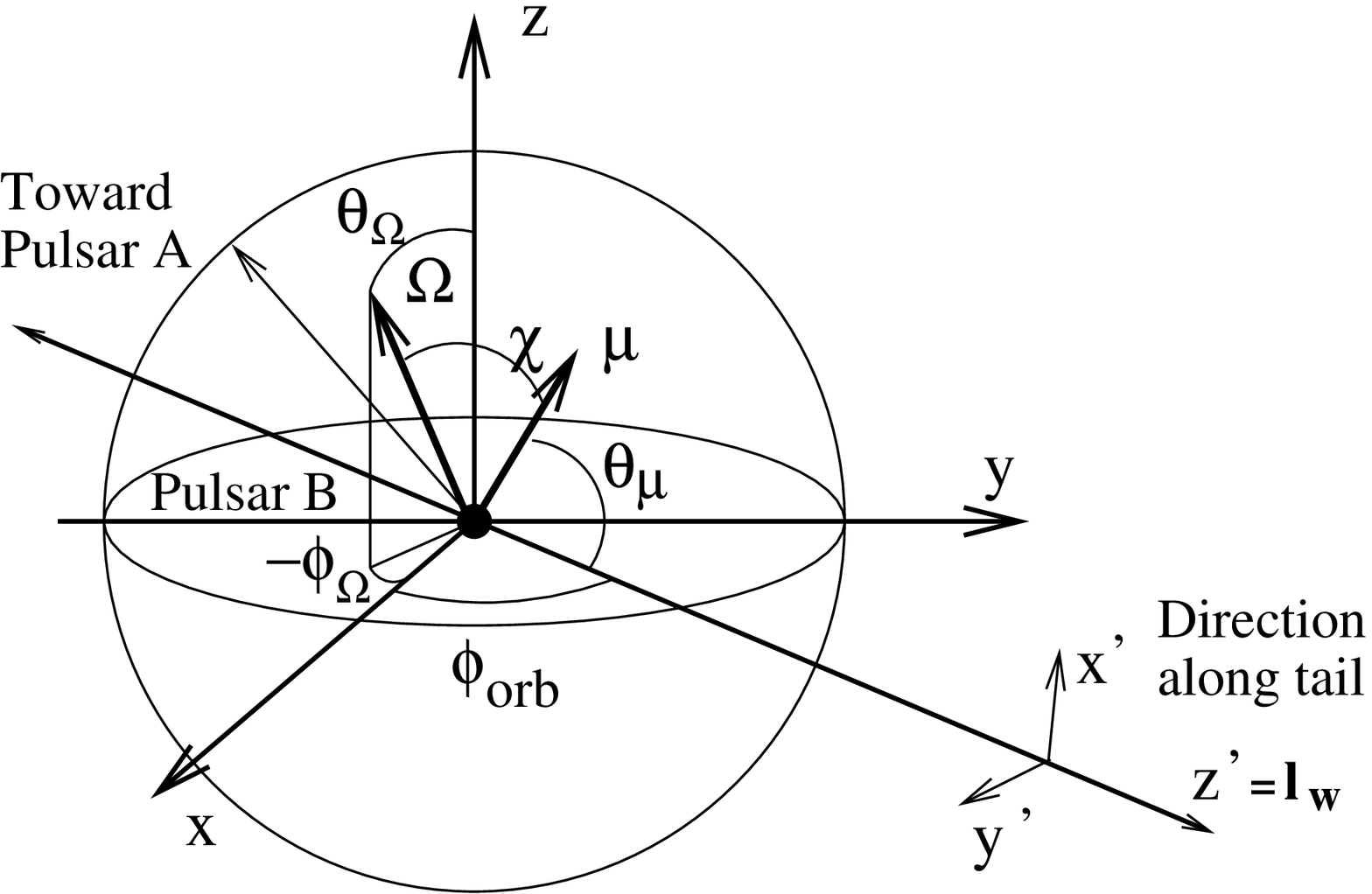}
\caption{Geometry of the model. Observer is located
at $x \rightarrow \infty$. 
The spin axis of pulsar B is defined by angles 
$\theta_\Omega$ and $\phi_\Omega$.  The direction of the magnetic moment 
$\vec\mu$
makes angle $\chi$ with the  spin axis and rotates around $\Omega$
every 2.77 s.
A line connecting two pulsars lies
in the x-y plane and makes angle $\phi_{orb}$ with $x$ axis.
(Note that we define orbital phase with respect
to the point of superior conjunction
and not with respect to ascending node. Two definitions differ by $270^\circ$.)
Tail coordinates $x',y',z'$ are also shown with vector $ {\vec \mu}$ in the 
$x'-z'$ plane. Angle $\theta_\mu$ is between $z'$ and $ {\vec\mu}$. 
}
\label{geom-orb}
\end{figure}

Let us choose a system of Cartesian  coordinates in the tail of B magnetosphere,
 centered on pulsar B so that 
$z'$ axis is along the line connecting pulsar A and B and axis
$x'$ is in the ${\vec\mu} -z'$ plane, where ${\vec\mu}$ is 
 magnetic moment of B,
see Fig. \ref{geom-orb}.  Let the 
 magnetic dipole be inclined at angle $\theta_\mu$ to axis $z'$.
The undistorted dipole Euler potentials are
\ba &&
\alpha_0 = R_{NS} \mu {( -x' \sin \theta_\mu + z' \cos \theta_\mu)^2+y^{\prime 2} \over
\sqrt{x^{\prime 2}+y^{\prime 2}+z^{\prime 2}} }
\nn &&
\beta_0=  R_{NS} \arctan {y^{\prime}\over -x^{\prime}\sin\theta_\mu +z^{\prime}\cos \theta_\mu}
\label{Euler}
\ea
Stretching transformations along $z^{\prime}$ axis are defined by potential
$f(z')$, so  that new Euler potentials are
expressed in terms of dipolar ones: $\{\alpha,\beta \}=\{\alpha_0(z'=f(z')), \beta_0(z=f(z'))\}$.
A degree of stretching depends on $f'(z')$.  In modeling 
of the Earth magnetosphere  function $f(z')$ is chosen to fit satellite data. 
Since we are interested in the distortions at one particular   location
 (assuming that radio emission is generated in  a narrow range of radii),
we choose $f'(z')=C=$const $<1$, similar to \cite{Voigt} model of 
Earth magnetosphere.  Thus, $C$ measures the distortion of magnetosphere at the emission radius.

Substituting $z' \rightarrow C z'$ in Eq. (\ref{Euler}) we find 
stretched magnetic fields in coordinates $\{x',y',z'\}$
\ba && 
{\bf B} ={ r_0^3 \mu \over \tilde{r'}^5} 
\left\{ 3 Cx^{\prime}z \cos \theta_\mu+ ( 2 C^2 x^{\prime 2} -y^{\prime 2} -z^{\prime 2}) \sin \theta_\mu,
\right.
\nn &&
3y^{\prime}C (z^{\prime}\cos \theta_\mu + Cx^{\prime}\sin \theta_\mu,
\nn &&
\left.
 C  \left(
3 C x'z' \sin \theta_\mu+ (2 z^{\prime 2} - C^2 x^{\prime 2} -y^{\prime 2}) \cos \theta_\mu \right)
\right\}
\ea
where $\tilde{r'}=\sqrt{ C^2 x^{\prime 2}+ y^{\prime 2} +z^{\prime 2}}$.
Integrating
along a magnetic field 
line in the $y'=0$ plane  we find equation for  magnetic surfaces:
\be
{r \over r_0} =
{ \left( C \cos  \theta_\mu \sin \theta' - \sin \theta _\mu \cos  \theta'\right)^2
\over \left(
\cos^2 \theta' + C^2 \sin^2  \theta' \right)^{3/2} }
\label{r}
\ee
where $r=\sqrt{x^{\prime 2}+z^{\prime 2}}$ and $\theta' = \arcsin x'/r'$ are 
polar coordinates 
aligned with $z^{\prime}$ and $r_0$ is a parameter related to maximum
extension of a field line.
From Eq. (\ref{r}) we find that polar field lines in the tail of  
stretched magnetosphere
are  defined by 
\be
\tan \theta_{p} = {1\over C} \tan \theta_\mu
\label{thetap}
\ee
This provides an  estimate of the  deviation of  polar field lines from the
direction of  magnetic dipole.
   Qualitatively, polar field lines are pushed toward
$z'$ axis. 
The method of field line stretching is only approximate and has  limited
applications. Its main drawback is that it offsets force balance, so that
there is a non-vanishing Lorentz force in the new configuration.
In addition, since the stretching method has been devised for magnetotail,
it's not clear how well it reproduces a  structure of the inner magnetosphere
(in  original \cite{Voigt} 
model the  stretching method is applied to 
tailward 
distances larger than approximately half the stand-off distance). 

\section{Orbital modulation of B}

Let us introduce another  Cartesian system of coordinates $x,y,z$ centered
on pulsar B, so that its orbital plane  lies in
 the $x-y$ plane (see Fig. \ref{geom-orb}).
The spin axis of pulsar B is inclined at an angle
$\theta_\Omega$ to the orbital normal, and at angle $\phi_\Omega$
with respect to the $x-z$ plane.
The magnetic moment of pulsar B has a magnitude
$\mu$,  is
inclined at an angle $\chi$ with respect to ${\bf\Omega}$ 
and executes a circular motion with phase $\phi_{rot}=\Omega t$, so that
$\phi_{rot}=0$ corresponds to the magnetic moment in the ${\bf\Omega}- x $ plane.
At a given orbital position, the unit vector along
the direction of  pulsar A wind
is approximately ${\bf l}_w= \{ \cos \phi, \sin \phi,0 \}$. 

In the observer frame the components of the unit vector $\hat {\bf \mu}(t)$ 
along instantaneous magnetic moment $\vec\mu(t)$ are
\be
\hat\mu_x  = \hat\mu_x^\Omega;
\;\;\;\;
\hat\mu_y  = \cos\theta_\Omega\,\hat\mu_y^\Omega +
\sin\theta_\Omega\,\hat\mu_z^\Omega;
\;\;\;\;
\hat\mu_z  = \cos\theta_\Omega\,\hat\mu_z^\Omega -
\sin\theta_\Omega\,\hat\mu_y^\Omega.
\ee
where $\hat\mu$ are coordinates  in a
 system aligned with ${\bf\Omega}$,
\be 
\hat\mu_x^{\Omega} = \sin\chi \cos(\phi_{rot} );
\;\;\;\;
\hat\mu_y^{\Omega} = \sin\chi \sin(\phi_{rot});
\;\;\;\;
\hat\mu_z^{\Omega} = \cos\chi.
\ee

Using Eq. (\ref{thetap}) we can find the direction of 
magnetic polar field lines
in the distorted magnetosphere:
\be
 {\bf s}_p = { \hat {\bf \mu} -(1-C) \cos \theta_{\mu} {\bf l}_w 
\over \sqrt{ 1+ (3-4C+C^2) \cos ^2 \theta_{\mu}   } }
\label{s}
\ee
where 
$\cos \theta_{\mu} = |\hat {\bf \mu}^0  \cdot {\bf l}_w |$ is the absolute value
of cosine of the angle between the direction of the wind and the direction
of the magnetic moment.

To estimate an influence of the orbit-dependent  magnetic field 
distortion on observed radio emission, 
we assume that   emission is generated near the polar field line in a region
with half opening  angle of $\sim
2 $ degrees, in accordance with the width of  pulsar B emission profile.
 Thus,  if the magnetic polar field line deviates from
the line of sight by more that $2 $ degrees, 
 we expect  emission of B to be weak.
The trajectory that magnetic polar field line makes on the sky depends 
on many parameters. To limit available phase space we use the results of 
modeling of pulsar A eclipse
\citep{lt05}, which imply that $\theta_\Omega \sim 60^\circ$ and 
$\chi \sim 75^\circ$. In total, we have to fit at least 6 parameters: $\phi_\Omega, \theta_\Omega,
\chi$, magnetospheric radius, impact parameter and distortion
coefficient $C$. Using results of \cite{lt05},
 the two principal parameters   that remain to be determined
are $\phi_\Omega$ and $C$. 
For a given set of $\theta_\Omega , \phi_\Omega, \chi $ and $C$,
the  direction of the  polar field line executes a non-circular  curve 
on the sky. An observer will see emission 
when for some values of   pulsar B rotation phase $\phi_{rot} $ and
orbital phase $\phi_{orb}$ the polar field line points
 within two degrees of  the line of sight.

\begin{figure}[h]
\includegraphics[width=0.9\linewidth]{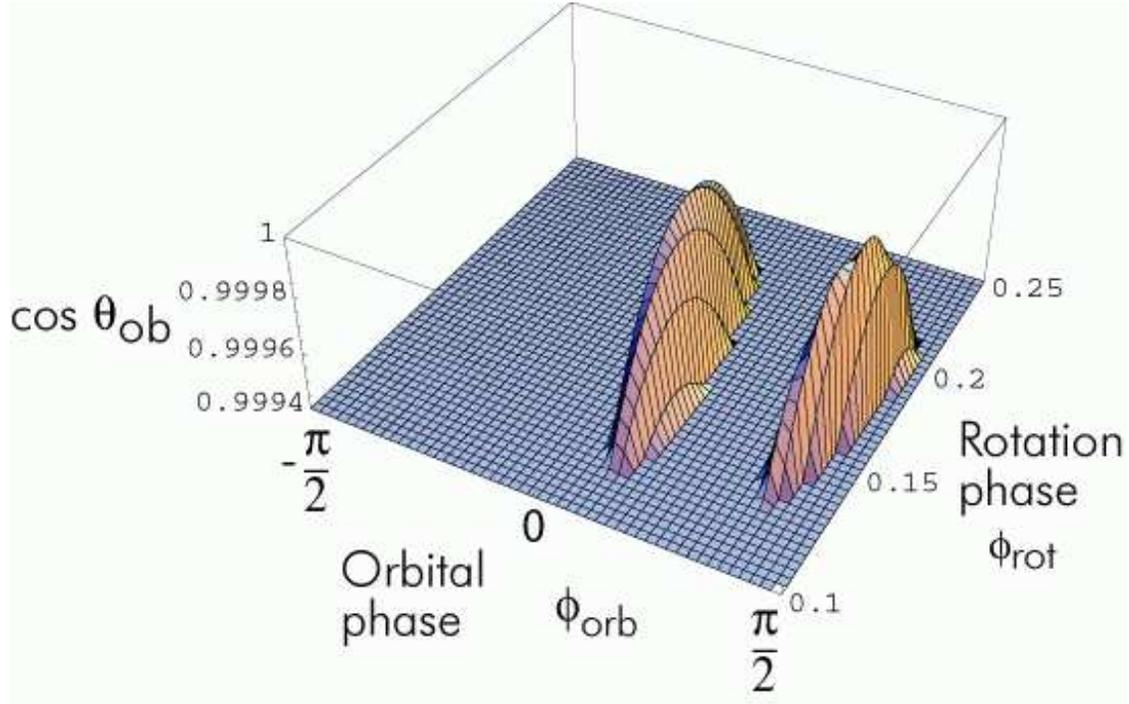}
\caption{Cosine of the angle $\theta_{ob}$
between the line of sight and the direction
of the polar field line in a distorted magnetosphere as a function of 
rotational phase $\phi_{rot}= \Omega t$ and orbital phase  
$-\pi/2 <\phi_{orb}< \pi/2$
measured from the superior conjunction. 
Parameters of the model are $\theta_\Omega = 60^\circ$, 
$\phi_\Omega = 67.5 ^\circ$, $\chi =73.6^\circ$, $C=0.7$.
Emission   is visible if $\theta_{ob}< 2 ^\circ$ corresponding to 
$\cos \theta_{ob} > .9994$. This happens at two orbital phases
$\phi_{orb}=0.03...0.43$ ($1.7^\circ ... 24^\circ$) 
and $\phi_{orb}=1.09...1.45$ ($62^\circ..83 ^\circ$). }
\label{costheta}
\end{figure}

\begin{figure}[h]
\includegraphics[width=0.9\linewidth]{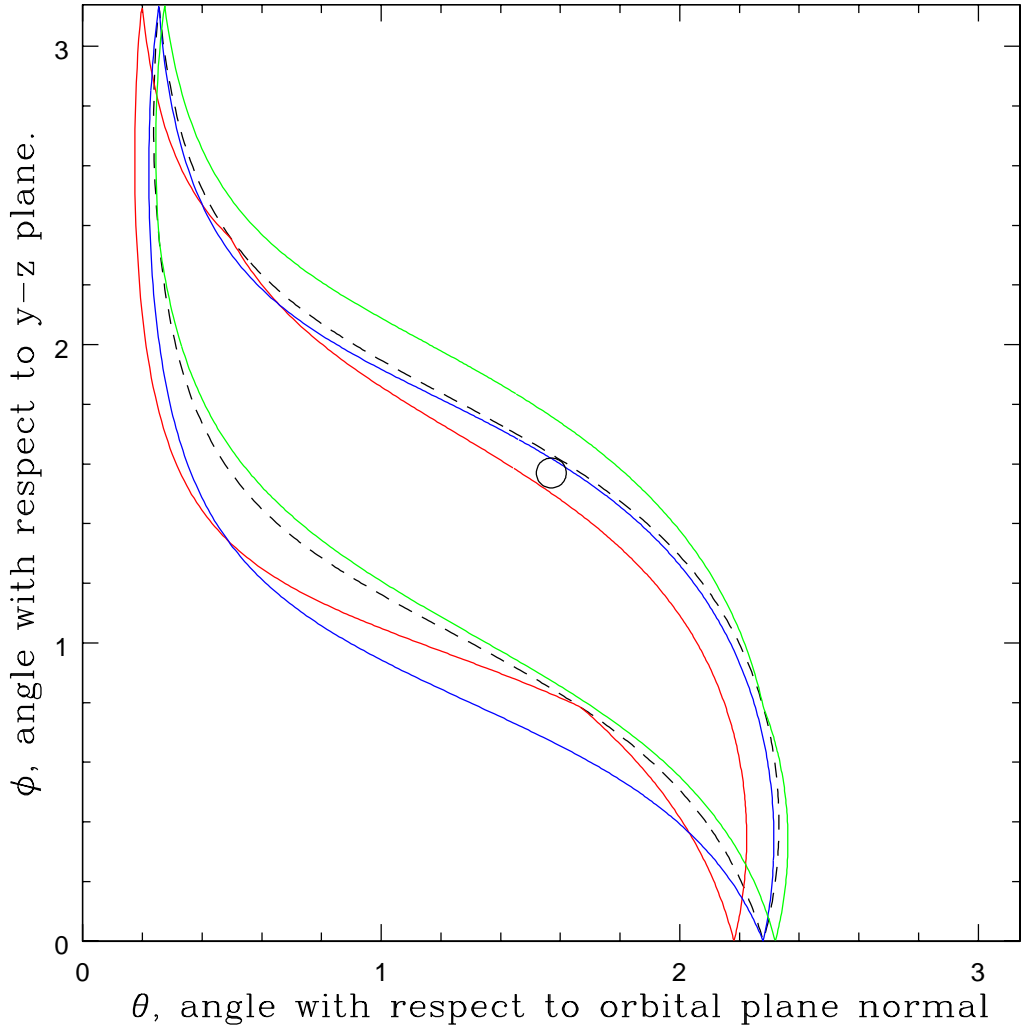}
\caption{Trajectory of the polar field line on the sky for
$\phi_\Omega = \pi/8=22.5 ^\circ$, $C=.7$.
Black dashed line: undistorted dipole, green solid is  at orbital phase
$\phi_{orb}=-\pi/4$ (misses the observer), blue line  
is at orbital phase $\phi_{orb}=0$
and red line is at  orbital phase $\phi_{orb}=\pi/4$. The location of  observer
is denoted by the circle. Upper and lower sets of curves correspond to two
magnetic poles.}
\label{path}
\end{figure}

\begin{figure}[h]
\includegraphics[width=0.9\linewidth]{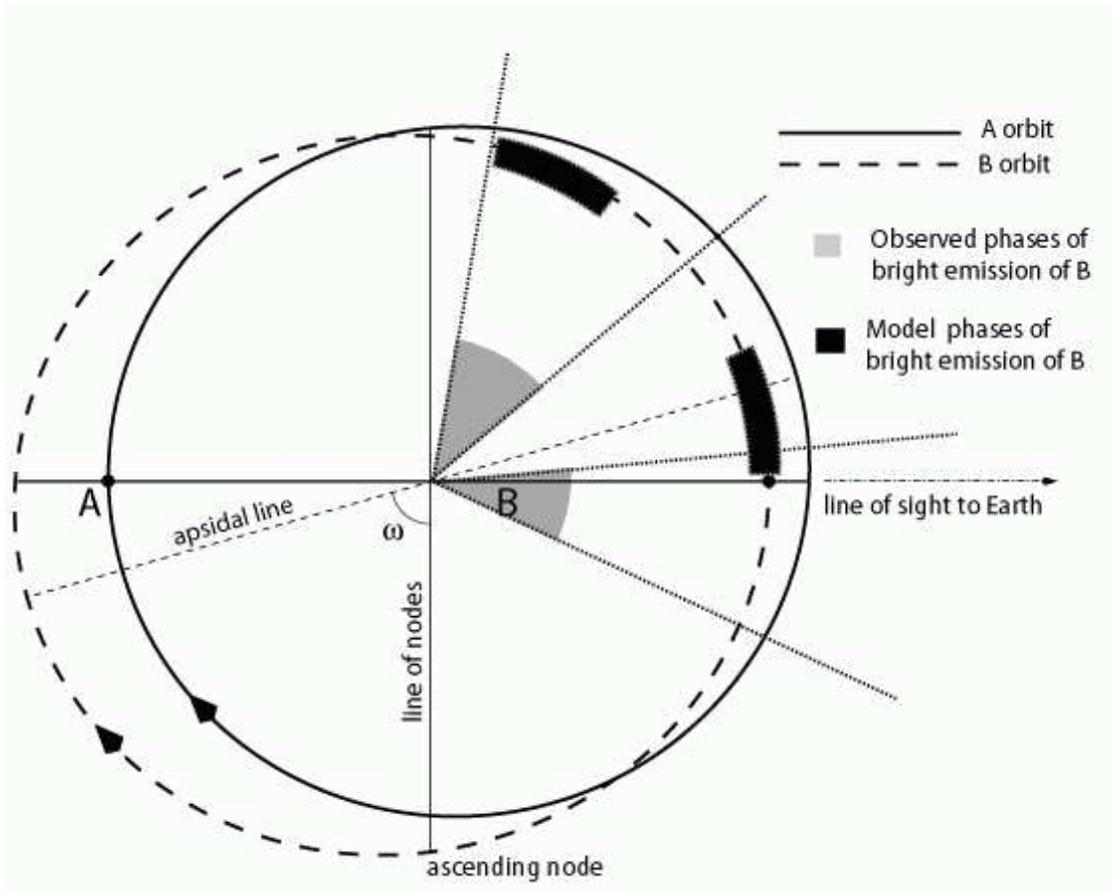}
\caption{Configuration of the system, after \protect\cite{lyne}. Light 
shades segments indicate
 orbital phases where B emission is strongest, dark  regions  indicate  the
location of emission in the best fit model. 
}
\label{emissionphase}
\end{figure}

Searching through parameter space we were trying to reproduce
two emission windows located in the tail of B magnetosphere.
After a number of trials, 
our best fit parameters are  $\phi_\Omega = - 67.5^\circ$ and $C=0.7$.
To illustrate the fit,
in Fig. \ref{costheta} we plot a  value of 
$\cos \theta_{ob}=   {\bf s}_p \cdot \hat {\bf x}$ as a 
function of orbital phase $\phi_{orb}$ and rotational phase $\phi_{rot}$.
Only points located above the line $\cos 2^\circ =.9994$ produce a pulse
of radio emission. (We restrict ourselves to
$-\pi/2 < \phi_{orb} < \pi/2$, corresponding to the  tail
pointing towards the observer.) 
Clearly, in the tail of  magnetosphere  the polar field
lines are pointing toward an observer at 
two orbital phases separated by approximately $70^\circ$,
see Fig. \ref{emissionphase}. 
One
of the phases,  nearly coincident with the conjunction, is  located at orbital phase
$\phi_{orb}=1.7^\circ ... 24^\circ$ and the second is located  at 
$\phi_{orb}=62^\circ...83 ^\circ$.
Considering the simplicity of the model and the fact that we had to make a
multi-parameter fit, the agreement with observations is impressive.
In order to  illustrate the effect of field distortion,
in Fig. \ref{path} we plot  a trajectory of 
magnetic polar lines on the sky, folded over $\pi$ radians to show both
poles. Emission is seen only at points approaching the direction to the 
observer within two degrees.  

Though we were able to obtains a satisfactory fit,
 we cannot guarantee that there is no other island in the
phase space of 6 parameters that also satisfies these criteria. 
In addition, there are intrinsic ambiguities in the model, related 
 to prograde versus retrograde rotation of pulsar B 
 and to which magnetic pole is seen by the observer.

\section{Discussions and Predictions}

We have constructed a simple  model of  orbital variations of 
pulsar B emission.  Small distortions of the inner magnetosphere
of B by  pulsar A wind change direction of the polar field lines, pushing
radiative beam of B towards the line of sight at two orbital phases.
 The main implication of the model is that B is always intrinsically bright.
 The model reproduces fairly well  absolute orbital location of 
bright emission windows, their width $\sim 20^\circ$ and separation
$\sim 70^\circ$.
It also naturally explains
different profiles at two emission windows, since at different
orbital phases the line of sight
crosses the emission region along different paths.

In this paper we considered only the nightside  magnetosphere.
The stretching method does not produce a realistic
structure of the dayside \citep{Stern94}. Qualitatively, we expect that
on the dayside
polar field lines will  also be  shifted from the direction of the magnetic pole
and  will be pushed out of the line of sight, producing a dip in the 
light curve of B close to the inferior conjunction. 
Since  emission beam is very narrow and  distortions can be considerable, 
it is fairly easy for the beam of B to miss an observer.
A more detailed model of magnetosphere is deferred to a subsequent  paper.

Our suggestion that pulsar B is always intrinsically bright is consistent with
its spin-down and  radio  energetics. First, assuming that an extension
of the last open field line is determined by the size of magnetosphere and that typical 
current density flowing on the open field lines is of the order
of the Goldreich-Julian current density, the total potential
over open field lines (a quantity that is usually related to efficiency
of pair production, \cite[\eg][]{arons79}) is independent of $R_m$:
$\Phi_{tot} \sim  \sqrt{L_{SD}/ 4 \pi c} $,
 where $L_{SD}=1.6 \times 10^{30} $ 
 erg/s is the spindown luminosity of B
 \cite{lyne}. This  ranks it as the 20th smallest (but not exceptional) out of nearly 1500
 pulsars with measured spindown luminosities (see 
 www.atnf.csiro.au/research/pulsar/psrcat). 
Secondly, its peak luminosity of $\sim 3$ mJy at 820MHz (Ransom, priv. comm.)
is typical for isolated pulsars with similar properties.

\begin{figure}[h]
\includegraphics[width=0.9\linewidth]{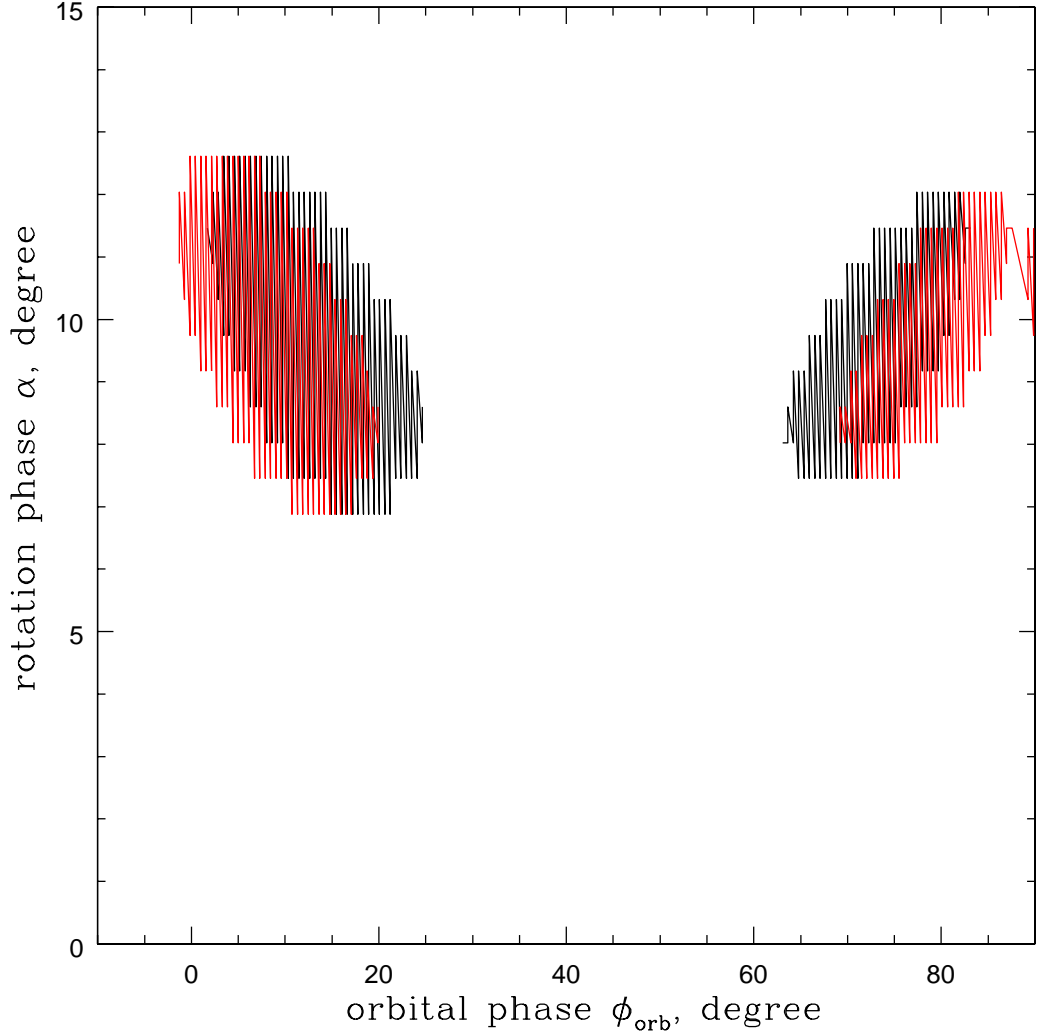}
\caption{
Dependence of the rotation phase of emission on the orbital position 
and evolution of the emission pattern due to changes of  $\phi_\Omega $.
B radio emission is seen at Earth at two shaded areas. 
At different orbital phases peak of emission occurs at 
different rotation phases. Typical variations in
 emission phase are of the order of the  profile width. The corresponding drift
  of arrival times for pulsar B is $\sim 15 $ ms.
Region in black is the current best fit $\phi_\Omega = - 67.5^\circ$, 
region on red is for $\phi_\Omega = - 68.5^\circ$.}
\label{alphaphi1}
\end{figure}

There is a number of predictions of the model. First,
at different orbital phases the center pulse corresponds to 
somewhat different rotational phases. Near $\phi_{orb}=1.7^\circ$, the 
center pulse is at $\phi_{rot} \sim 11.5 ^\circ$, decreasing to 
$\phi_{rot} \sim 8^\circ$
at  $\phi_{orb}=24^\circ$, 
remaining the same at  $\phi_{orb}=62.5^\circ$ and increasing
back to  $ \phi_{rot}\sim 11.5 ^\circ$ 
at $\phi_{orb}=83.1^\circ$, see Fig. \ref{alphaphi1}.
Thus, one expects a drift of the profile as a function of an orbital phase.
Particular values for the drift angle  and rate 
are strongly dependent on the precise parameters
of the model, but  the type of evolution is generic: during a visible phase
the profile drifts approximately by its width.
If averaged over a bright
emission window, the  drift of the emission phase may be interpreted as a
large timing noise of B.

A drift of the emission phase as a function of  orbital position may 
have already been observed. \cite{ransom}  reported a
systematic change in arrival times of B pulses by 10-20 ms. 
This is consistent with the prediction of the model, since a change in 
arrive phase by 2 degrees of rotation phase of B  corresponds to $\sim 15$ ms. 

In our model, the weak emission of B observed throughout the orbit has a 
different  origin than the bright emission. For example, it can be generated
in a much wider cone, akin to interpulse  emission (bridges) observed 
in regular pulsars.

 Our second prediction is related to a secular 
evolution of pulsar B emission properties.
 The emission beam of B  is fairly narrow, so that small changes  in the
orientation of rotation axis of B may induce large apparent changes 
in the profile. One possibility for the change in the direction
of the spin is a geodetic precession of B, which should happen
on a relatively short time scale $\sim 70 $ yrs \citep{lyne}. The 
geodetic precession will affect
mostly the angle  $\phi_\Omega$. From our modeling we find that changes of 
$\phi_\Omega$ by as little as $\sim 1^\circ$ strongly affect observed B 
profile (see Fig. \ref{alphaphi1}). 
 Thus, we expect that profile of B may change on a times scale of less than a  year.
 According to the model, average profile width  remains approximately constant.
Changes of the orbital phases of emission are not  accompanied by  
substantial changes in the  emission
phase. (Note that if 
the emission geometry is non-trivial, \eg elliptic instead of 
circular, one does expect  changes in the  emission
phase.)
Since at different epochs line of sight passes through different emission regions
one may expect variations in pulse intensity.
Thus, using different slices  taken at different epochs one can construct
a detailed  map of the emission region. This should prove  a valuable method
in constraining pulsar radio emission mechanisms.

\begin{figure}[h]
\includegraphics[width=0.95\linewidth]{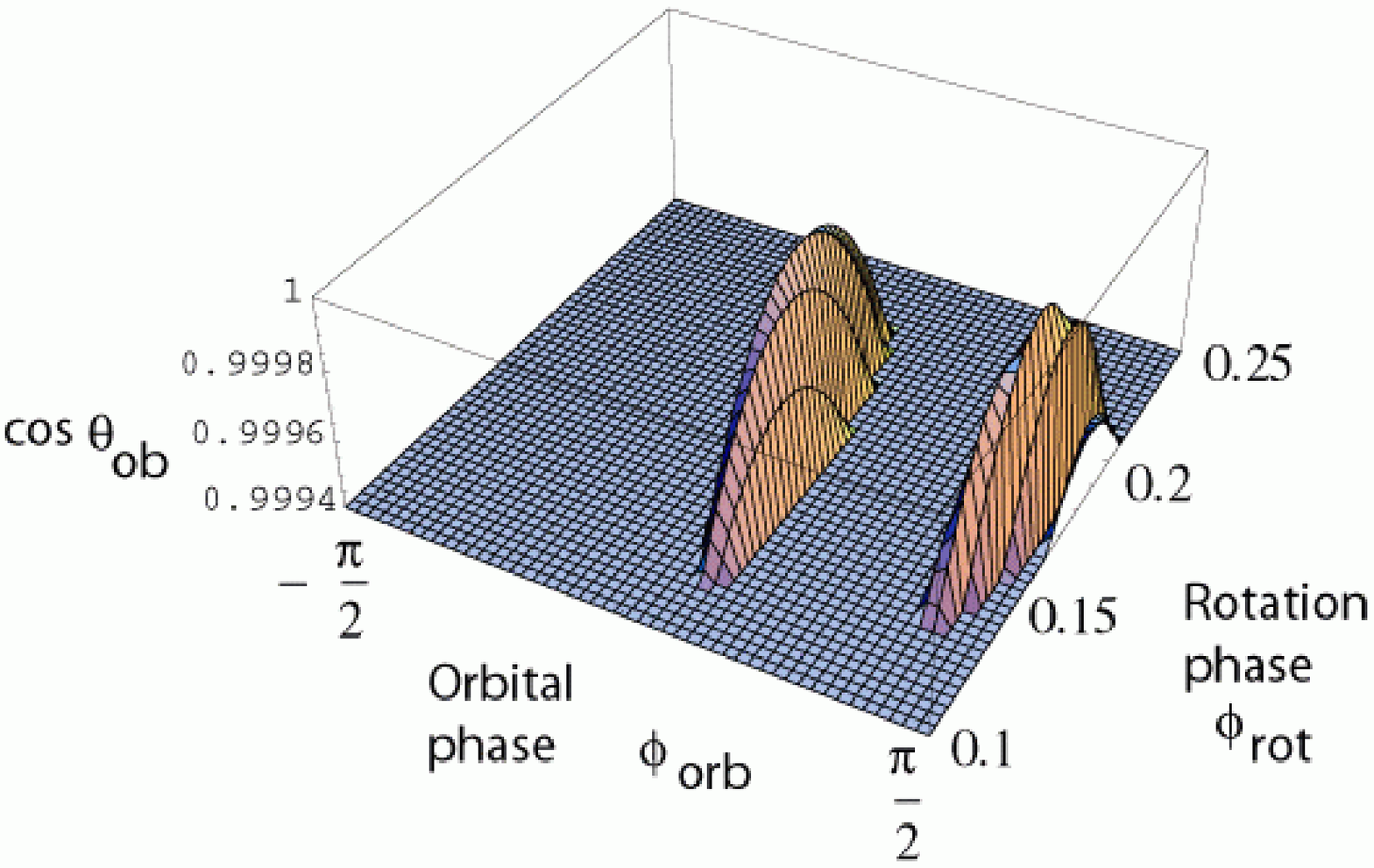}
\includegraphics[width=0.95\linewidth]{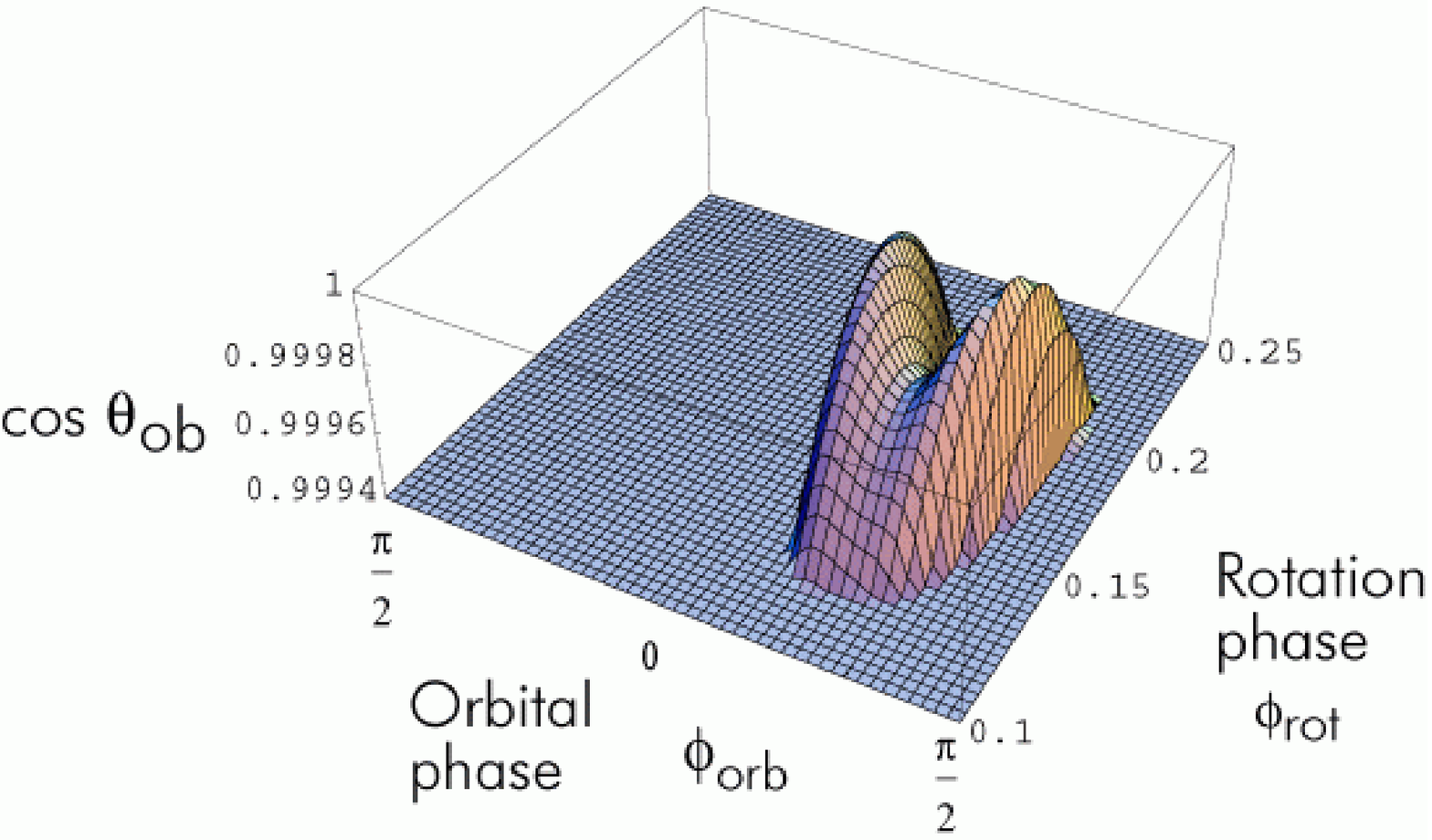}
\caption{Secular changes in the position of  the emission regions due to geodetic drift:
(a)  $\phi_\Omega = -68.5^\circ$, (b)   $\phi_\Omega =- 65^\circ$
(current best fit value is $\phi_\Omega =- 67.5^\circ$). 
In case (a) the orbital separation between the  two bright phases 
increases, in case (b) it decreases, so that two phases merge in one.
Except for $\phi_\Omega$,  parameters are the same as in Fig. \ref{costheta}.
}
\label{costhetadr}
\end{figure}

A longer evolution of the profile
 cannot be predicted unambiguously since we do not know the direction of the 
drift.
Two possibilities include increasing or decreasing $|\phi_\Omega + 90^\circ|$, 
see Fig. \ref{costhetadr}. 
In one case, two emission regions 
get closer together merging in one, while in the other case they separate
and a new one appears approximately at a mirror reflection of the first, at $\phi_{orb} \sim - 
 90^\circ$. 
 \footnote{According to the model,
 we are not likely to lose pulsar B in the coming year, yet the model
 is not sufficiently detailed to guarantee it.}
We would like to stress again that exact details of the secular evolution are hard to predict 
 using this very simple model.

 For the stretching coefficient $C=0.7$, the 
relative deformation of the magnetosphere at the emission radius is $\sim 30\%$. 
This is a fairly large distortion, which 
 favors  large emission altitudes,
$R_{em}\geq  10^8$ cm. Large emission altitudes in isolated pulsars
have been previously suggested by  \cite{lbm98}.
A more precise modeling of B \Bf  may reduce required distortion and thus
allow somewhat smaller $R_{em}$. Still, 
near stellar surface distortions are expected to be tiny, so that 
 the model requires high emission altitude.

The success of this simple model is somewhat surprising, given the fact that in all we had to fit
many parameters with a required precision of $2$ degrees. 
Qualitatively, the reason for the success
of this model, as well as that of \cite{lt05}, is that, to the first order,
magnetic field  is  well approximated by dipolar structure.
Given the simplicity of the
model, some of the parameters may  not be well determined: small
variations in parameters may induce relatively large observed  changes. 

A number of 
 effects  may complicate the picture.
First of all, 
a non-trivial 
geometry of the emission region, \eg
elliptical instead of circular, may increase
 the quality of the fit.  The fact that B profiles in the two emission windows
 are different implies that emission geometry is indeed non-circular. A double hump 
 profile in one of the windows also points to a more complicated emission geometry.
Secondly, non-spherical A wind will produce distortions 
dependent on orbital phase. Also, if reconnection between 
wind and magnetospheric field lines is important, the structure of  magnetosphere may depend 
on the direction of the wind \Bf. 
We plan to address these issues in a subsequent paper.
Clearly, a detailed modeling of B magnetosphere is required to further finesse
the model.

When the paper has been mostly completed we learned the results of 
\cite{burgay05}, who found secular changes in B profiles
in general
agreement with predictions of the model.

\begin{acknowledgements}
I would like to thank Maura McLaughlin and
Ingrid Stairs for numerous enlightening discussions and comments on the manuscript.
I also thank Gerry Atkinson, Marta Burgay,  Joeri van Leeuwen and 
Scott Ransom.
\end{acknowledgements}

{}

\end{document}